# COINS CHANGE LEADERS – LESSONS LEARNED FROM A DISTRIBUTED COURSE

Peter A. Gloor

MIT CCI
5 Cambridge Center
Cambridge MA 02138
pgloor@mit.edu

Maria Paasivaara

Aalto University
Espoo, Finland
maria.paasivaara@aalto.fi

## ABSTRACT

In this paper we analyze the communication network of 50 students from five universities in three countries participating in a joint course on Collaborative Innovation Networks (COINs). Students formed ten teams. Interaction variables calculated from the e-mail archive of individual team members predict the level of creativity of the team. Oscillating leadership, where members switch between central and peripheral roles is the best predictor of creativity, it is complemented by the variance in the amount of sending or receiving information, and by answering quickly, and positive language. We verify our automatically generated creativity metrics with interviews.

## INTRODUCTION

Researchers disagree if breakthrough innovation relies on the lonely creative genius (Simonton 2013, Murray 2004), or is done by the creative group (Salazar et al. 2012, Sawyer 2007). In our research we investigate the communication behavior of creative people and groups, addressing the apparent dichotomies identified by Czikszentmihalyi (1996) of famously creative people being both full or energy as well as at rest, being playful but disciplined, and being both introvert and extrovert. Over the last ten years we have been studying success factors of innovation through the lens of social network analysis (SNA) developing the framework of Collaborative Innovation Networks (COINs) (Gloor 2006). In particular we measure how centralization of a network changes; using the metrics of group betweenness centrality (Freeman 1977) and contribution index (Gloor et al. 2003). Contribution index measures how active as senders and recipients of information network members are.

In a series of experiments we have compared creativity and performance with networking patterns in e-mail networks of students (Gloor et al. 2007), open source programmers (Kidane & Gloor 2007) and bank employees (Gloor et al. 2010), and face-to-face networks of students and programmers (Gloor et al. 2012), and Jazz musicians (Gloor et al. 2013) using sociometric badges (Olguin 2007) that measure direct interpersonal interaction. In this project we apply the same approach – comparing communication structure with team creativity and performance – to multinational student teams.

## PROJECT SETUP

In a seminar 50 students from five universities (MIT Cambridge, MA; Savannah College of Art and Design (SCAD) GA; Aalto University, Helsinki, Finland; University of Cologne, Germany and University of Bamberg, Germany worked together for five months in multinational virtual project teams as COINs (Collaborative Innovation Networks). They formed ten teams ranging in size from three to six students from at least two locations, working on a project related to social media and social network analysis. Students were asked to send all their project-related e-mail communication to a dummy e-mailbox. This allowed us to construct a virtual mirror of ongoing communication within and between teams. At the end of the course, each team presented their results to their classmates in a virtual meeting. Each of the ten presentations was ranked by the students and the instructors in three categories: presentation quality, content quality, and creativity on a scale from 1 to 5 where 5 was the lowest and 1 the highest. Comparing the virtual mirror of communication with the peer and instructor ratings permitted us to identify the communication patterns leading to the most highly-ranked work output.

## HYPOTHESES

Rotating leadership indicated by oscillating betweenness centrality was first observed among Eclipse open source developers (Kidane & Gloor 2007), and subsequently verified with 16 members of the marketing team of a bank in a face-to-face network (Fischbach et al. 2009), as well as among nurses in a hospital (Olguin et al. 2009) where we compared face-to-face interaction with individual personality characteristics measured by the Neo-FFI

(McCrae et al. 2005). Besides individual creativity measured as openness in the Neo-FFI we also rated group creativity though peer and management/instructor assessment, based on the premise that experts know creativity when they see it (Amabile 1983).

We therefore formulate our first hypothesis

*1. the higher the oscillation in group betweenness centrality over time of a team is, the more creative it is.*

We also found that teams where the workload over time is shared evenly in terms of sending and receiving e-mails among team members, are more productive. It is measured through the variance in contribution index (Gloor et al. 2007), see also caption of figure 2. We therefore conjecture:

*2. The lower the average weighted variance in contribution index is, the higher is team performance.*

## RESULTS

Using the ten dummy e-mailboxes of the project teams, we constructed a combined social network (figure 1). In the group network in figure 1 the different teams can be clearly recognized. The communication of each team is shown in a different color, usually team members are clustered together as a COIN, with external collaborators and other students being in more peripheral positions. The network map was created using the email communication among the students and instructors over the entire course. Each dot is an actor and the lines between them represent one or more emails.

Analyzing the contribution index (figure 2) shows that members of the same team tend to show similar behavior regarding the ratio of e-mails sent to e-mails received. Clusters of dots of the same color are members of the same group, overall we find that higher-performing teams tend to communicate more actively, with more similar send/receive ratios.

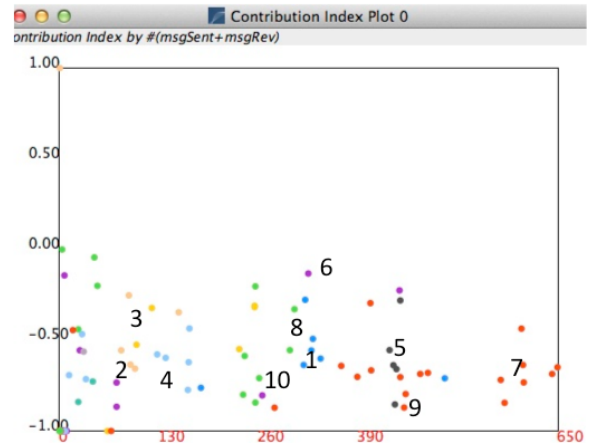

*Figure 2. The Y-axis is the Contribution Index, which ranges from -1 to +1. "1" means that that person only sends emails; "0" means a perfect balance of sending and receiving emails; and "-1" means that a person only receives emails. The X-axis is the count of the number of emails.*

The temporal social surface (figure 3) (Gloor 2005) indicates creativity, as there is a relatively large group of high-betweenness class members which is constantly changing over time, in earlier work we found this to be a reliable predictor of creativity.

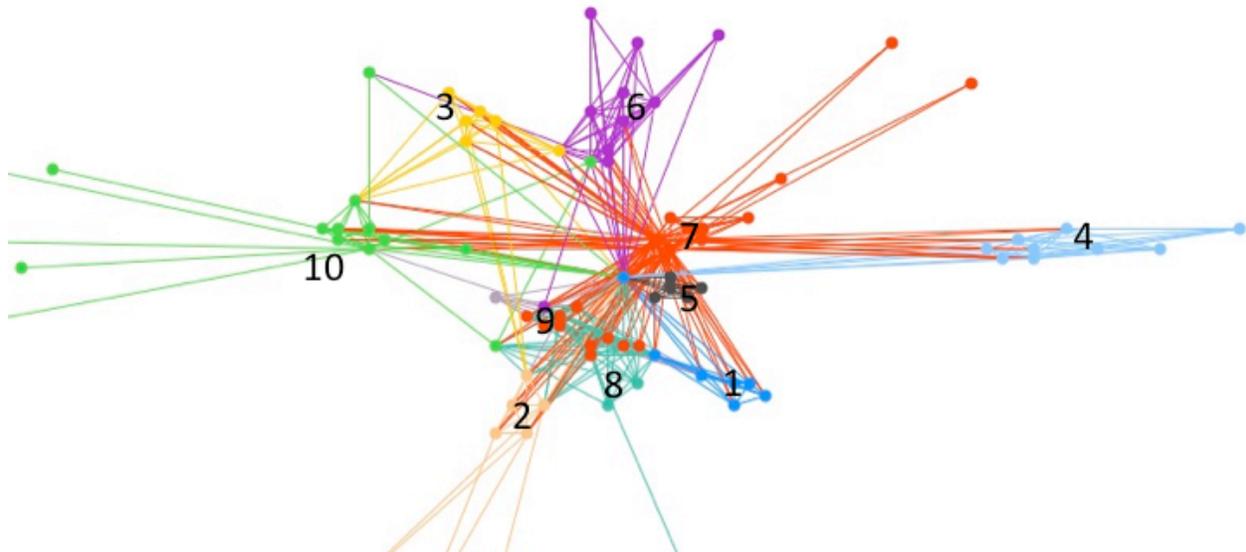

*Figure 1. Team network*

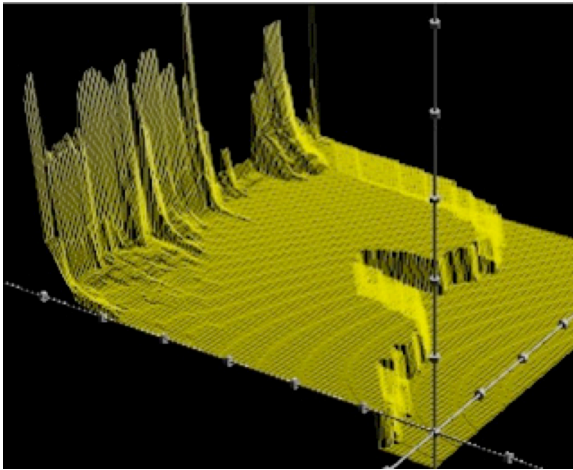

*Figure 3. The Temporal Social Surface is composed of three items: The Y-Axis are the students, the X-axis is time in days and the Z-Axis is the betweenness centrality value. The back plane that rises and falls represents a set of students who rotated taking the lead in team communication.*

The 6 snapshots of the 10 teams' communication networks over the 5 months (figure 4) show Tuckman's four phases in the life of a team (Tuckman 1965): forming, storming, norming, and performing. We see how the main instructor in the uppermost picture at right is most central, but how then teams start connecting in the middle row, and how they then huddle together team-by-team to focus on their work in the bottom-most pictures.

The group betweenness centrality curve as well as the absolute number of messages sent and received shown in figure 5 illustrate the higher traffic in the forming, storming, and norming phase, followed by the lower traffic in the second performing phase.

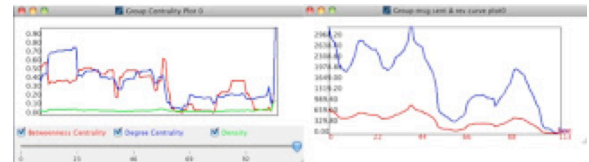

*Figure 5. Group betweenness and degree centrality, and density over time (left) and number of messages sent and received (right)*

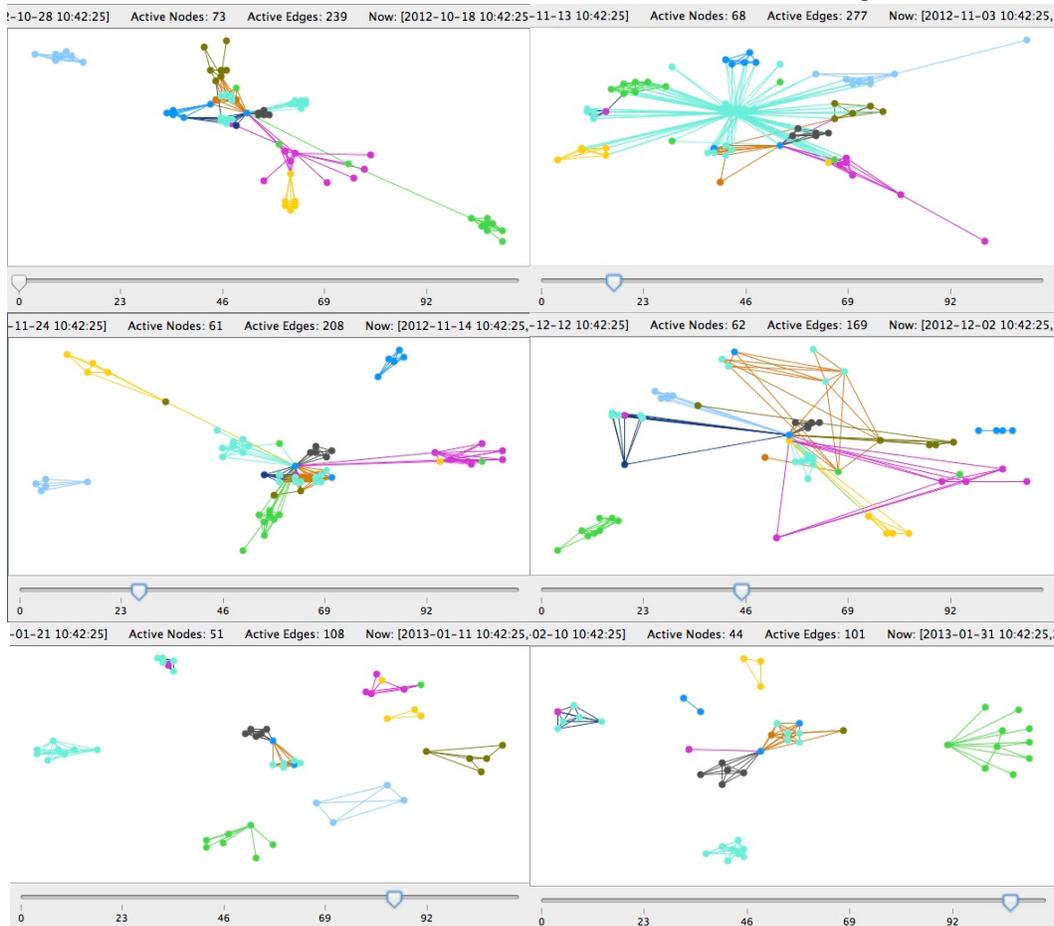

*Figure 4. Tuckman's four phases of group development in the e-mail network*

The sentiment curve (figure 6) illustrates the same phenomenon, with higher emotionality (defined as the sum of positivity and negativity) in the forming and storming phase in the first half of the course. The X-axis is always days in figures 5 and 6. Sentiment is calculated by a simple bag-of-words approach, where counts of positive and negative mood words are normalized by document size.

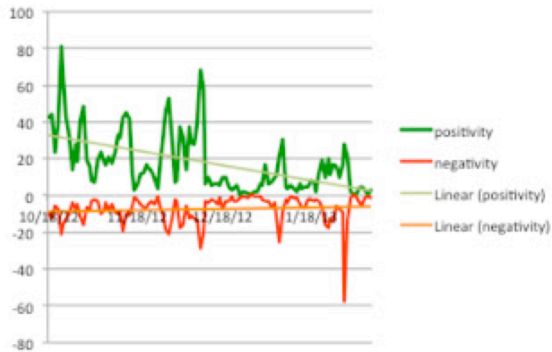

*Figure 6. Positive and negative sentiment over entire course duration*

As the correlations in table 2 illustrate, the instructor rating of creativity on a scale from 1 to 5 where 1 is best and 5 is worst (each of the instructors at each of the five participating locations ranked the 10 presentations) (table 1) correlates highly (0.83**) with oscillation in betweenness centrality. The team rated the most creative had 82 oscillations, i.e. handovers in leadership, compared to the lowest rated team with only 1 oscillation.

| team # | Creativity (low is better) | num of bc oscillations | ART Norm (min) | Pos_Sent | AWVCI | gbc_strong_tie | group_dc | Msg Recvd | num_actors |
|---|---|---|---|---|---|---|---|---|---|
| 1 | 2.4 | 5 | 88.67 | 2.696 | 0.022 | 0.127 | 0.133 | 1172 | 7 |
| 2 | 2.6 | 8 | 311.00 | 0.961 | 0.398 | 0.556 | 0.607 | 350 | 9 |
| 3 | 3 | 4 | 382.65 | 0.558 | 0.552 | 0.382 | 0.196 | 667 | 9 |
| 4 | 2.8 | 1 | 79.00 | 0.211 | 0.068 | 0.310 | 0.222 | 686 | 10 |
| 5 | 1.4 | 82 | 78.25 | 2.087 | 0.051 | 0.160 | 0.000 | 2133 | 6 |
| 6 | 2.6 | 13 | 201.00 | 0.131 | 0.118 | 0.466 | 0.407 | 852 | 15 |
| 7 | 2.2 | 23 | 135.04 | 2.193 | 0.058 | 0.525 | 0.956 | 3804 | 64 |
| 8 | 2 | 20 | 67.27 | 2.644 | 0.050 | 0.660 | 0.509 | 1975 | 20 |
| 9 | 1.6 | 26 | 74.44 | 2.150 | 0.060 | 0.576 | 0.419 | 2431 | 18 |
| 10 | 2.6 | 1 | 343.38 | 1.566 | 0.279 | 0.894 | 0.738 | 1361 | 17 |

*Table 1: Basic Data of 10 student COINs course teams (ART=Average Response Time; AWVCI-Average Weighted Variance in Contribution Index)*

It was also found that the more positive the content of the e-mails sent is, the higher is the team's creativity. This might be a consequence of the international team composition, which opens up room for cultural misunderstanding. The best way to overcome such obstacles is to give positive reinforcements wherever appropriate.

There is also significant correlation between average response time to e-mails, and creativity. The lower the e-mail response time, the higher is team creativity, i.e. the more responsive team members are to each other's inquiries, the higher the creative quality of the work output.

We found that the larger the variance in contribution index is, the lower is the team's creativity (p=0.06). This means that a few people send significantly more than the rest, i.e. there might be some free riders in the project team, leading to a lower quality result.

Finally the sheer number of messages exchanged also correlates with creativity, indicating that a small team working on a creative task can never communicate enough.

**STUDENT INTERVIEW RESULTS**

After the course had ended, we interviewed six students, each from a different project team and asked them to share their collaboration experiences, tell about their learning during the course, as well as give feedback on the course.

As the most important learning during the course the students mentioned collaboration in an international multi-disciplinary team, where the different members have different skillsets and thus, as a group, they can complement each other's skills. For example a combination of design, business and programming skills in a group was highly appreciated.

*"Actually it [multi-disciplinarity] worked to our advantage. (...) so it was really nice getting to work with different people at different stages in their career, as well as their educational backgrounds (...) it was a pretty dynamic group of backgrounds, which brought a lot to the table. (...) what was nice is that individually we had all these different elements that helped push our project one step further. (...) All of us were kind of able to bring something together to*

|  |  | Num bc oscillations | ART Norm | Pos Sent | AWV CI | bc_strong_tie | group_dc | num_actors | Msg Recvd |
|---|---|---|---|---|---|---|---|---|---|
| Creativity low is better | Pearson Correlation<br>Sig. (2-tailed)<br>N | -.830**<br>.003<br>10 | .656*<br>.039<br>10 | -.684*<br>.029<br>10 | .612<br>.060<br>10 | .109<br>.764<br>10 | .118<br>.745<br>10 | -.121<br>.740<br>10 | -.652*<br>.041<br>10 |
| Bc coscillations | Pearson Correlation<br>Sig. (2-tailed)<br>N |  | -.439<br>.204<br>10 | .370<br>.293<br>10 | -.376<br>.284<br>10 | -.355<br>.314<br>10 | -.328<br>.355<br>10 | -.001<br>.997<br>10 | .467<br>.173<br>10 |
| ART Norm min | Pearson Correlation<br>Sig. (2-tailed)<br>N |  |  | -.495<br>.146<br>10 | .930**<br>.000<br>10 | .408<br>.242<br>10 | .259<br>.470<br>10 | -.131<br>.718<br>10 | -.475<br>.166<br>10 |
| posSent | Pearson Correlation<br>Sig. (2-tailed)<br>N |  |  |  | -.500<br>.141<br>10 | .005<br>.989<br>10 | .111<br>.761<br>10 | .277<br>.439<br>10 | .648*<br>.043<br>10 |
| awvci | Pearson Correlation<br>Sig. (2-tailed)<br>N |  |  |  |  | .251<br>.485<br>10 | .071<br>.845<br>10 | -.251<br>.485<br>10 | -.534<br>.112<br>10 |
| gbc_strong_tie | Pearson Correlation<br>Sig. (2-tailed)<br>N |  |  |  |  |  | .776**<br>.008<br>10 | .317<br>.373<br>10 | .129<br>.723<br>10 |
| group_dc | Pearson Correlation<br>Sig. (2-tailed)<br>N |  |  |  |  |  |  | .768**<br>.010<br>10 | .431<br>.214<br>10 |
| num_actors | Pearson Correlation<br>Sig. (2-tailed)<br>N |  |  |  |  |  |  |  | .801**<br>.005<br>10 |

*Table 2. Correlations between creativity and SNA metrics. \*\*. Correlation is significant at the 0.01 level (2-tailed). \*. Correlation is significant at the 0.05 level (2-tailed).*

*help push towards that direction and get a conclusion."* – A US Student

Students had noticed that collaborating with people having different disciplinary backgrounds was more challenging than with their classmates from the same university, but at the same time also rewarding. Creating a common understanding in this kind of a distributed group with different backgrounds was mentioned as the biggest challenge. Some interviewed students even mentioned that everybody in their group had thought that they had understood what they should do, but they soon found out they all had understood it differently. Most groups spent quite a lot of time in the beginning of a project just to agree on a focus for their groupwork topic. Most topics were quite broad, thus each group needed to find a reasonably narrow scope. As could be seen from the team formation graphs (Figure 4), in this "focus-searching phase" most groups spent a reasonable amount of time in coming to agreement both through emails and in voice/videoconference meetings, before they could concentrate on performing the work.

*"When we started on this project, we were a little bit confused. (...) But it turned out quite well. (...) Once we got to a certain point, where we no longer felt that we were kind of swimming in a sea of information, we were able to divide and conquer."*
- A US student

*"Our topic was very open, not that focused (...), so I guess the problem was then (...) how to focus, and what to really work on and decide what we are doing. (...) I think we were quite often jumping a little bit back and forth and saying ´Oh maybe we should also look at that and that…' (...) Overall, as a group (...) we had a little bit different opinions on how much we should explore certain areas, but usually also in our Skype meetings it was not that we would end up into fighting or anything, it was a kind of constructive discussion (...) we didn't have like bad disagreements. (...) As a team we worked together, and it didn't lead into any disharmony in the team (...) Finding clear focus, that was the biggest challenge. (...) everybody kind of tried out different things and then at some point we focused…"*
- A Finnish student

Besides email communication, all groups had voice/videoconference meetings over the Internet, some weekly, some bi-weekly. These meetings served their purpose very well: first as brainstorming sessions in discussing on the topic and searching for focus and later on in following up on progress and dividing the work tasks. Each team had an assigned mentor, either one of the teachers, or e.g., an external "customer" for the team. Mentors participated in part of the teams' meetings, which was highly appreciated by the teams. Our interviewees especially commented that they found it inspiring when the mentor did not bring them ready-made answers, but only suggested things they could think about and that way just guided them to the right direction.

*"We got a lot of input from [mentor]. He had expectations about the goal of our project, but he gave us a lot of freedom on how we can do it. He usually answered to our emails with some directions or hints on 'that could be the way to go', but he never said 'do it this way', he said 'you should look at this, and maybe go to that direction', which was pretty good I think. We always knew we were on the right track or not on the right track and that helped a lot I think."*
- A German student

Regarding communication, interviews confirmed what we could see from the email networks: the teams collaborated as true distributed teams, instead of forming, e.g., site-specific sub-teams. Some teams had formed pairs intentionally to accomplish some tasks, e.g., one team formed three sub-teams to research three different sub-topics, and after accomplishing that task, chose one of the sub-topics and continued again working as a whole team.

*"What we usually did was to break things up into three teams with two people tackling each item."*
– A US student

We were happy to notice that when pairs for pair work were formed, they were based on complementary skills and areas of interest, not on location. Thus, many of the pairs were formed across the sites.

The interviews also confirmed what we could see from communication networks: the projects did not have named project managers, but persons in more active leadership roles often took turns, as could be seen from the following quotation:

*"…manager role (…) we decided to actually to keep it more flexible (….) so it was a little bit of taking turns, that different people were maybe driving it a bit more at different times, which overall also went, I would say, pretty well. Of course there were people that were not that active and others who were active…"*
– A Finnish student

## CONCLUSIONS

The conclusions for high-functioning teams are threefold: First and foremost, it pays to pass the baton frequently, the more leadership rotates among team members, the more creative their output will be. Secondly, speed of communication matters; the faster instructor and students engage in a dialogue, the higher the quality if their work output. Finally, we found that it is best to use honest language: praise when praise is due, but also say when something is not ok.